\documentclass[useAMS,usenatbib]{mn2e}
\usepackage{epsfig,amsmath,natbib,amssymb,relsize,graphicx}
\usepackage{color}
\usepackage{ifpdf}

\def\refjnl#1{{\rm#1}}
\def\aj{\refjnl{AJ}}                   
\def\apj{\refjnl{ApJ}}                 
\def\apjl{\refjnl{ApJ}}                
\def\apjs{\refjnl{ApJS}}               
\def\aap{\refjnl{A\&A}}                
\def\mnras{\refjnl{MNRAS}}             
\def\ssr{\refjnl{Space~Sci.~Rev.}}     
\def\nat{\refjnl{Nature}}              

\def\be{\begin{equation}} 
\def\ee{\end{equation}}

\def\msun{{\Msun}}

\def\HH{${\rm {H_2}}\,\,$}

\def\fesc{f_{\esc}}

\def\gsim{\lower.5ex\hbox{\gtsima}} 
\def\lsim{\lower.5ex\hbox{\ltsima}} \def\gtsima{$\; \buildrel > \over 
\sim \;$} \def\ltsima{$\; \buildrel < \over \sim \;$} \def\prosima{$\; 
\buildrel \propto \over \sim \;$} \def\gsim{\lower.5ex\hbox{\gtsima}} 
\def\lsim{\lower.5ex\hbox{\ltsima}} 
\def\simgt{\lower.5ex\hbox{\gtsima}} 
\def\simlt{\lower.5ex\hbox{\ltsima}} 
\def\simpr{\lower.5ex\hbox{\prosima}} \def\la{\lsim} \def\ga{\gsim} 
  
\def\gtsima{$\; \buildrel > \over \sim \;$} 
\def\ltsima{$\; \buildrel < \over \sim \;$} 
\def\gsim{\lower.5ex\hbox{\gtsima}} 
\def\lsim{\lower.5ex\hbox{\ltsima}} 
\def\simgt{\lower.5ex\hbox{\gtsima}} 
\def\simlt{\lower.5ex\hbox{\ltsima}} 
\def\simpr{\lower.5ex\hbox{\prosima}} 
\def\la{\lsim} 
\def\ga{\gsim}

\def\fesc{f_{esc}}

\def\msun{\,{\rm \Msun}}

\def\E3{{\cal E}_{\rm g}^{III}}

\def\Msun{\rm M_\odot}

\def\Tvir{T_{vir}} 
\def\r12{r_{1/2}} 
\def\x12{x_{1/2}} 
\def\v12{v_{1/2}}

\def\mbh{m_{\bullet}}
\def\hs{\hspace{1mm}}
\def\jcrit{J_{\rm crit}}

\def\jlw{J_{\rm LW}}
\def\jbg{J_{\rm LW}^{\rm BG}}
 
\title[Conditions for direct collapse black hole formation]{Feedback-regulated Super Massive Black Hole Seed Formation} 
\author[Dijkstra et al.]
{Mark Dijkstra$^{1,2}$\thanks{E-mail:mark.dijkstra@astro.uio.no},Andrea Ferrara$^3$ \& Andrei Mesinger$^3$\\
$^{1}$Institute of Theoretical Astrophysics, University of Oslo, Postboks 1029, 0858 Oslo, Norway\\
$^2$Max Planck Institute for Astrophysics, Karl-Schwarzschild-Str. 1, 85741, Garching, Germany\\
$^{3}$ Scuola Normale Superiore, Piazza dei Cavalieri 7, I-56126 Pisa, Italy\\
}

\begin{document} 
 
\date{\today} 
 
\pagerange{\pageref{firstpage}--\pageref{lastpage}} \pubyear{2012} 
\voffset-.45in 
\maketitle 
 
\label{firstpage} 
\begin{abstract} 
The nature of the seeds of high-redshift supermassive black holes (SMBHs) is a key question in cosmology. Direct collapse black holes (DCBH) that form in pristine, atomic-line cooling  halos, illuminated by a Lyman-Werner (LW) UV flux exceeding a critical threshold  $\jcrit$ represent an attractive possibility. We investigate when and where these conditions are met during cosmic evolution.  For the LW intensity, $\jlw$, we account for departures from the background value in close proximity to star forming galaxies. For the pristine halo fraction, we account for both ({\it i}) supernova driven outflows, and ({\it ii}) the inherent pollution from progenitor halos. We estimate the abundance of DCBH formation sites, $n_{\rm DCBH}(z)$, and find that it increases with cosmic time from $n_{\rm DCBH}(z=20) \sim 10^{-12}-10^{-7}$ cMpc$^{-3}$ to $n_{\rm DCBH}(z=10) \sim 10^{-10}-10^{-5}$ cMpc$^{-3}$. Our analysis shows the possible importance of galactic winds, which can suppress the predicted $n_{\rm DCBH}$ by several orders of magnitude, and cause DCBH formation to preferentially occur around the UV-brightest ($M_{\rm UV} \sim -21\pm 1$) star forming galaxies. Our analysis further highlights the dependence of these predictions on ({\it i}) the escape fraction of LW photons, ({\it ii}) $\jcrit$, and ({\it iii}) the galactic outflow prescription.
 \end{abstract}


\section{Motivation}
\label{Mot}
The process by which astonishingly massive ($\mbh \approx 10^9 \Msun$) black holes came into existence by the time ($\simlt 1$ Gyr) at which they are now discovered (\citealt{Fan01a, Fan01, Fan06}; \citealt{Willott09, Willott10}; \citealt{Mortlock11}; \citealt{Venemans13}) in progressively larger numbers, is one of the most puzzling mysteries in cosmic evolution. The current paradigm 
implies that these Super Massive Black Holes (SMBH) have grown, starting from a much smaller seed, via matter accretion and, to a lesser extent by merging with other compact objects (\citealt{Volonteri03}; \citealt{Volonteri05}; \citealt{Lodato06}; \citealt{Volonteri03}; \citealt{Natarajan11}; \citealt{Tanaka09}; \citealt{diMatteo08}; \citealt{Li07}). This hypothesis has recently become less tenable as several authors pointed out a number of related difficulties.  

The growth rate of the SMBH can be written as 
\begin{equation}
\frac{d\ln m}{dt} = \frac{1-\epsilon}{\epsilon}\frac{1}{t_E}  
\end{equation}
where $t_E = [4\pi G\mu m_p/\sigma_e c]^{-1}= 0.45$ Gyr is the Eddington time, and $\epsilon$ denotes the radiative efficiency. In order to grow a SMBH of mass $\mbh(t)$ at cosmic time $t(z)$ corresponding to redshift $z$, we need an initial BH seed of mass
\begin{equation}
m_0 = \mbh(t) \exp\left[{-\frac{1-\epsilon}{\epsilon}\frac{t(z)}{t_E}}\right].  
\end{equation}
Then, assembling the SMBH mass ($\mbh=2\times 10^9 M_\odot$) deduced for the most distant quasar ULAS J1120+0641 at $z=7.085$ \citep{Mortlock11} when $t(z) = 0.77$ Gyr, requires $\ln (m_0/M_\odot) > 21.4- 1.71(1-\epsilon)/\epsilon$.
For the usually assumed value of $\epsilon = 0.1$, this translates into $m_0 > 400 M_\odot$. Such value is uncomfortably large when compared to the most recent estimates of the mass of first stars, which now converge towards values $\ll 100 M_\odot$ \citep{Greif11,Stacey12,Hosokawa12a,Hirano14}. 

Long before these problems were realized, proposals for the production of more massive ($m_0 \approx 10^{4-6} \Msun$) seeds were
made (\citealt{Loeb94}; \citealt{Eisenstein95}), which have now developed into more complete scenarios (\citealt{Begelman06}; \citealt{Shang10}; \citealt{Johnson12}; \citealt{Regan09}; \citealt{Bonoli12}; \citealt{Latifa}; \citealt{Latifb}; \citealt{LatifSMS}; \citealt{Regan14}). This channel invokes the formation of massive black hole seeds in environments where, for reasons explained in the following, gas gravitational collapse proceeds at very sustained rates, $\dot M_g =\simgt 0.1 -1\Msun$ yr$^{-1}$ ( i.e. about 100 times larger than for standard metal-free star formation). These objects are often dubbed as ``direct collapse black holes'' (DCBH) to distinguish them from the smaller seeds of stellar origin discussed above. Where can these super-accreting environments be found? The most promising candidates are dark matter halos with virial temperature $\Tvir \simgt 10^4$ K. In these halos the primordial gas radiatively cools via collisional excitation of the hydrogen $1s \rightarrow 2p$ transition followed by a emission of a Ly$\alpha$ photon. Given the strong temperature sensitivity of such process, the gas collapses almost isothermally, $1+d\ln T/d\ln \rho \equiv \gamma \approx 1$, thermostating the temperature at $T\approx 8000$ K. Under these conditions, gas fragmentation into sub-clumps is almost completely inhibited (\citealt{Schneider02}; \citealt{Li03}; \citealt{Omukai05}; \citealt{Omukai08}; \citealt{Cazaux09}) and collapse proceeds to very high densities unimpeded. 

While atomic cooling keeps the gas on the isothermal track, if H$_2$, heavy elements or dust are present they can strongly decrease $T$, and thus induce fragmentation. Hence, a key point for the mechanism to work is that the collapsing halo is (a) metal-free, and (b)  exposed\footnote{\citet{IO12} have shown that the presence of a strong photodissociating background is not required when shock heating by cold accretion flows keeps the gas temperature at $T\sim 10^4$ K while the gas cloud contracts until it reaches a threshold density of $n\gsim 10^4$ cm$^{-3}$. Beyond this density threshold collisional dissociation keeps the gas $H_2$ free, and the gas remains at $T\sim 6000$ K (also see Fernandez et al. 2014). Conditions for this `UV-free' collapse may occur more frequently when baryonic streaming motions are accounted for \citep{Tanaka14}. More recently however, \citet{Visbal14} have shown that accretion shock heating to $T_{\rm vir}$ only occurs at densities up to a few orders of magnitude lower than $n\sim 10^4$ cm$^{-3}$ with hydrodynamical simulations, and trace this back to the entropy of the IGM when it decouples from the CMB. They conclude that a strong LW flux is required for DCBH formation.} to a sufficiently high external soft ($0.7 \simlt h\nu/\mathrm{eV} \simlt 13.6)$  UV radiation field to photo-dissociate the H$_2$ (or the catalyzer H$^-$) via the two-step Solomon process (\citealt[][]{Omukai01}). Recent works have shown that if the UV intensity, $J_\nu=J_{21} \times 10^{-21}$erg s$^{-1}$cm$^{-2}$Hz$^{-1}$\,sr$^{-1}$, is larger than a critical value, $\jcrit$, that then the abundance of \HH molecules is strongly depressed and that the cooling time always remains longer than the freefall time. 

The precise value of $\jcrit$ depends on the details of the UV spectral shape (\citealt{Shang10}; \citealt{Wolcott-Green12}); for a $T=10^4$ K blackbody spectrum, \citet{Shang10} showed that $\jcrit$ lies in the range $\jcrit=30-300$, while for a $T=10^5$ K blackbody spectrum we expect $\jcrit=10^3$ \citep[][]{Wolcott-Green12}. In this paper, we use \texttt{Starburst99} to generate spectra for star forming galaxies. These spectra are typically harder than that of a $T=10^4$ K blackbody, but softer than that of a $T=10^5$ K blackbody (Wolcott-Green et al. in prep). Thus far, there exists no published values for $\jcrit$ for this kind of `intermediate', more realistic spectrum in the literature (but see Wolcott-Green et al. in prep). We pick $\jcrit=300$ for our fiducial model, which is intermediate between the $\jcrit$ for the two different black-body spectra. We study the impact of choosing a lower $\jcrit$ as part of our analysis.

The LW-background was likely less than $\jcrit$: \cite{Petri12} have shown that the requirement that reionization was completed by $z=6$ with a ionizing photon/baryon ratio ${n_\gamma}/{n_b}=10$ (\citealt{Mitra12a}) corresponds to a LW intensity 
\begin{equation}
\jbg(z) =  0.14 \fesc^{-1} (\Omega_b h^2)(1+z)^3 \left(\frac{n_\gamma}{n_b}\right) \approx  10 \fesc^{-1} ,
\end{equation}  
where $\fesc$ is the mean escape fraction of ionizing photons from galaxies. Current studies (\citealt{Inoue06}; \citealt{Ferrara12}; (\citealt{Finkelstein}, \citealt{Kuhlen12}; \citealt{Becker13}) indicate that $\fesc$ increases from $f_{\rm esc}\sim 0.01-0.2$ at $z<8$ to about $f_{\rm esc} \sim 0.4-0.6$ at higher redshift. Thus, this simple estimate alone suggests that a LW background intensity $\jbg < \jcrit$ is likely. This is further confirmed with estimates for $\jbg$ obtained from numerical simulations (\citealt{Ciardi00}; \citealt{Machacek01}; see also a review in \citealt{Ciardi05}). Various recent analyses have found comparable values in the range $\jbg\sim 20-40$ at $z=8-20$ \citep{Dijkstra08,Ahn,HF12,McQuinn12}. 

When $\jbg$ does not exceed the critical value, the UV background fluctuates due to clustering of sources and, to a lesser extent,
radiative transfer effects (Dijkstra et al. 2008). 
\cite{Agarwal12} recently used cosmological N-body simulations, combined with simple prescriptions for 
star formation to calculate the spatially-varying intensity of the UV radiation and identify pristine haloes in which DCBH can potentially
form. They find that $J_{21}$ can be up to $10^6$ times the spatially averaged background, thus resulting in a very large 
abundance ($\approx 10 ^{-2} \Msun$ Mpc$^{-3}$) of DCBH. This value is in striking contrast with other studies
(\citealt{Dijkstra08}; \citealt{Tanaka09}) who instead find a number density of potentially DCBH host halos of 
$\simlt 10 ^{-6} \Msun$ Mpc$^{-3}$. The origin of this difference is mostly related to the fact that D08 used a $\jcrit =10^3$ \citep[a value that was found in earlier work, see e.g.][]{Omukai01,BL03}, while \citet{Agarwal12} adopted $\jcrit=30$, which can largely explain these differences (see \S~\ref{Sum} for a more detailed comparison).

The picture gets even more complex if one also requires the halo gas to meet condition (a) above, i.e. to be metal free. The fraction of halos that are below the critical metallicity for fragmentation ($Z_{crit} = 10^{-5\pm 1} Z_\odot$, e.g. \citealt{Schneider06}) depends on the details of mechanical and chemical feedbacks (\citealt{Schneider06a}; \citealt{Tornatore07}; \citealt{Salvadori12}) and must be carefully evaluated. \cite{Dijkstra08} did not explicitly include metal enrichment in their model, whereas \cite{Agarwal12} only consider pollution of the collapsing halo from previous episodes of star formation within it. This might not be the major source of pollution compared to outflows from nearby galaxies (\citealt{Scannapieco02}) if the such object are to be located in highly clustered regions where $\jlw > \jcrit$. 

To conclude, DCBH formation requires the following conditions on the host halos: (a) virial temperature $>10^4$ K to ensure high accretion rates allowed by atomic cooling; (b) gas metallicity $Z < Z_{crit}$ to prevent fragmentation into clumps induced by heavy elements and dust cooling; (c) they must be exposed to a UV intensity $\jlw > \jcrit = 300$ to strongly depress \HH abundances. The main aim of this paper is to quantify the number density of potentially DCBH host halos as a function of time\footnote{Throughout the paper, we assume a flat Universe with cosmological parameters  given by the PLANCK13 \citep{Ade} best-fit values: $\Omega_m=0.3175$, $\Omega_{\Lambda} = 1 - \Omega_m=0.6825$, $\Omega_b h^2 = 0.022068$, and $h=0.6711$.  The  parameters defining the linear dark  matter power spectrum are $\sigma_8=0.8344$, $ n_s=0.9624$.}. Two features in this paper distinguish this work from previous analyses are: ({\it i}) we study metal pollution by galactic outflows\footnote{After this paper was submitted, a preprint by \citet{Ag14} appeared which also includes the impact of galactic winds on DCBH formation sites in their high-resolution hydrodynamical simulation. We compare our results explicitly in \S~\ref{Sum}.} inside regions in which the local LW-background value is elevated due to the presence of a nearby star forming galaxy, and ({\it ii}) we present a systematic study of the dependence of our results on various model parameters. Throughout, we denote the Lyman-Werner background with $\jbg(z)$. All LW flux-densities are given in units of $10^{-21}$ erg s$^{-1}$ cm$^{-2}$ Hz$^{-1}$ sr$^{-1}$. For brevity we have dropped the usual `$_{21}$' subscript.

The outline of this paper is as follows: We describe our prescriptions for including metal enrichment in \S~\ref{Metal}. In \S~\ref{sec:lwbg} we present calculations of the global LW background values, and quantify the fluctuations in the background. We compute the number density of putative DCBH formation sites, $n_{\rm DCBH}(z)$, as a function of redshift in \S~\ref{Formation}, before discussing the implications of our results in \S~\ref{Sum}.

\begin{table}
\caption{Summary of different LW-related symbols used in this paper.}  \centering.
\begin{tabular}{l l}
\hline\hline\\
symbol & description \\
\hline \\
  $\jlw$ & LW intensity impinging on a collapsing cloud.\\
  $\jbg$ & intensity of LW background \\
 $J_{\rm LW}^{1s}$ & LW intensity generated by a single source \\
    $J_{\rm crit}$ & minimum (`critical') intensity \\
    & required for DCBH formation.\\
    $\epsilon_{\rm LW}(z)$ & LW-volume emissivity at redshift $z$\\
    $\langle L_{\rm LW} (M)\rangle$ & {\it average} LW-luminosity density\\
    & assigned to dark matter of mass $M$\\
    $\sigma_{\rm LW}$ & standard deviation used in lognormal\\
    & dispersion assumed in $L_{\rm LW}-M$ relation\\
  \hline\hline
\end{tabular}
\label{table:symbols}
\end{table}


\section{Metal pollution}
\label{Metal}

\subsection{Metal Pollution from Galactic Outflows}

Massive stars that produce UV \& LW radiation also produce and eject metals at the end of their lives
as supernovae. Metals spread out to some radius $r_{\rm s}(M,t)$ and they will trigger fragmentation
of the gas in halos at $r < r_{\rm s}(M,t)$, thus quenching DCBH formation. Here, $M$ denotes the host dark 
matter halo mass. Metals are transported by galactic outflows, and to a rough
approximation the radius of the metal-enriched region increases according to   
\be 
r_{\rm s}(M,t) = \left(\frac{E_0 \nu M_\star}{m_p n}\right)^{1/5} t^{2/5},
\label{eq4}
\ee 
where the total stellar mass $M_\star=f_\star M\Omega_{\rm b}/\Omega_{\rm m}$, $E_0 = 1$ Bethe is the supernova explosion energy, $\nu = 0.01 M_\odot^{-1}$ is the number of supernovae per solar mass of stars formed and $n$ is the number density of the gas in which the explosion goes off. This approximation fits within a factor of 2 the more detailed results of \cite{Madau01}. Under these hypotheses, and substituting some numbers we obtain
\be 
r_{\rm kpc,s}(M,t) = 3\times 10^{-2} \left(\frac{M_\star}{M_\odot}\right) ^{1/5} n^{-1/5} t_6^{2/5}.
\label{eq5}
\ee For the density $n$ entering Eq. \ref{eq5}, we assume that the gas is $\Delta=60$ times denser than the mean IGM value at redshift $z$, i.e. $n\approx 60 \times \Omega_{\rm b}\rho_{\rm crit}(1+z)^3/m_{\rm p}=0.021[(1+z)/11]^{3}$ cm$^{-3}$. This value corresponds to the typical baryonic overdensity of halos at their virial radius for a Navarro-Frenk-White profile.  As the blast wave will initially expand in the halo $\Delta = 60-1000$ and then spend most of its evolutionary time in the IGM at mean density $\Delta \approx 1$, the previous assumption tends to underestimate the bubble size and hence the size of the polluted regions in which DCBH cannot form. 

\begin{figure*}
\includegraphics[width=75mm]{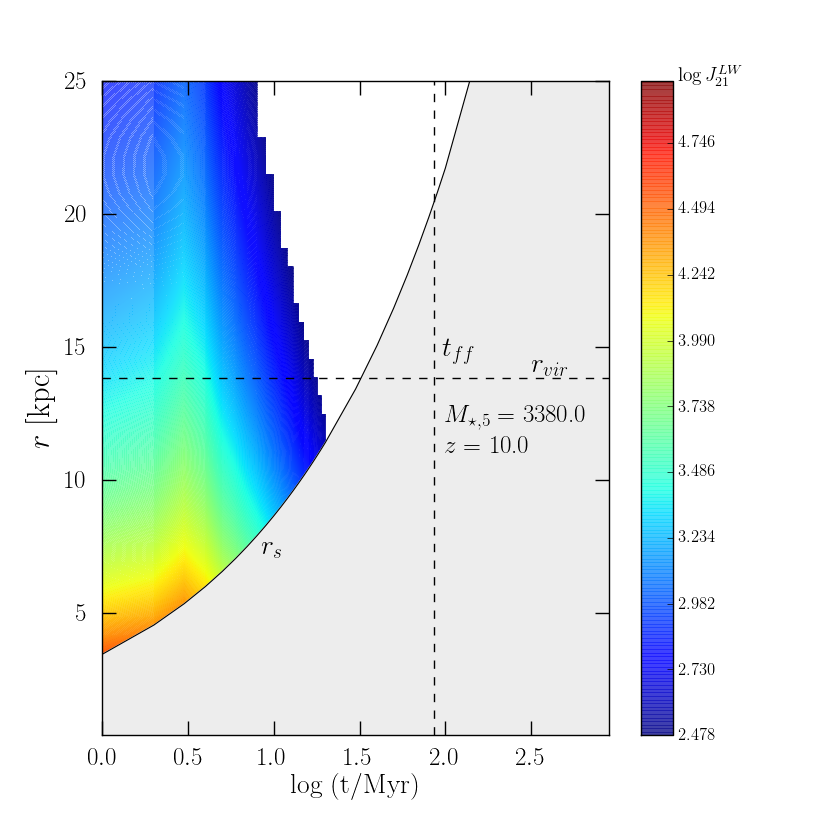}
\includegraphics[width=75mm]{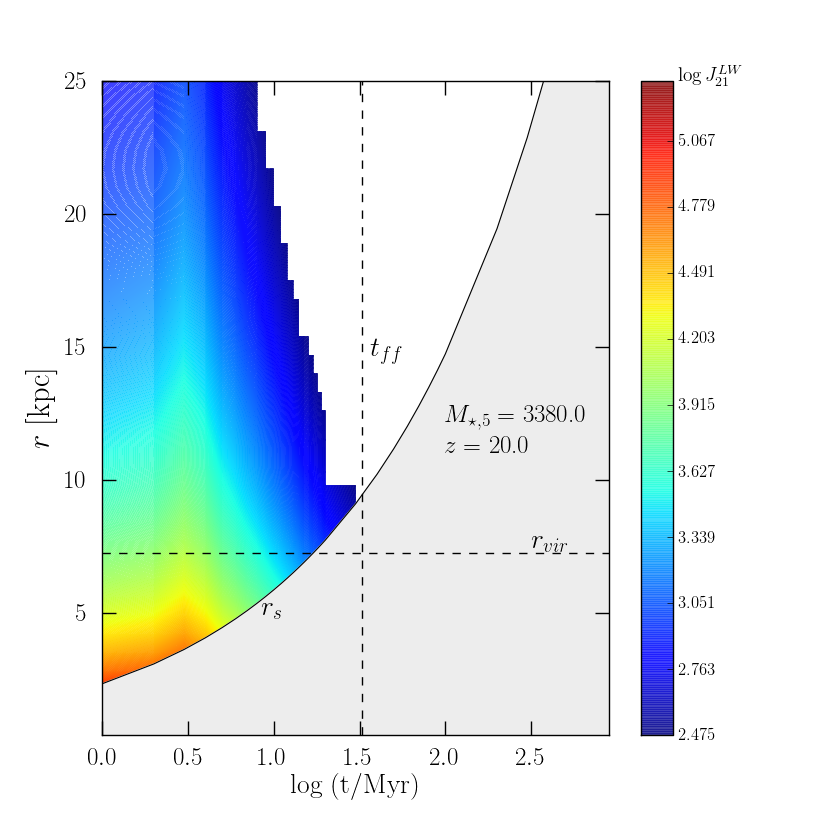}
\caption{Time evolution of radial extent of the region in which DCBH formation might occur in genetically pristine halos at $z=10$ ({\it left panel}) and $z=20$ ({\it right panel}) around a dark matter of mass $M\sim 4 \times 10^{10} \msun$. The colors of pixels in the $r-t$ plane indicate the value of $\jlw$ when $\jlw > \jcrit=300$. The {\it solid line} denotes the radius $r_{\rm s}$ polluted by the galactic outflow. The {\it grey region} at $r < r_{\rm s}$ is not suitable for DCBH formation. The {\it vertical} [{\it horizontal}] {\it dotted line} marks the free-fall time [virial radius] associated with the dark matter halo. In our models, we require a non-zero volume at $t=t_{\rm ff}$ as a condition for DCBH formation. In these example, this requirement is not met, which illustrates the possible importance of galactic winds.} 
\label{Evol}
\end{figure*}

Example of the time evolution of $r_{\rm s}$ at $z=10$ ($z=20$) is shown shown in the {\it left panel} ({\it right panel}) of Figure~\ref{Evol} for a galaxy with a total stellar of mass $M_\star=3\times 10^8 M_{\odot}$. For the models discussed in this paper, this corresponds to a dark matter halo mass of $M\sim 4 \times 10^{10} M_{\odot}$ (see \S~\ref{LWLum}). In this Figure the colour of a pixel in the $r-t$ plane shows the LW-flux, $\jlw$, where we only show pixels with $\jlw > \jcrit=300$. This Figure shows that the region exposed to $\jlw > \jcrit$ decreases with time, while the region sterilised by galactic winds (whose outer boundary is at $r=r_{\rm s}$) increases with time. 

\begin{figure}
\vbox{\centerline{\epsfig{file=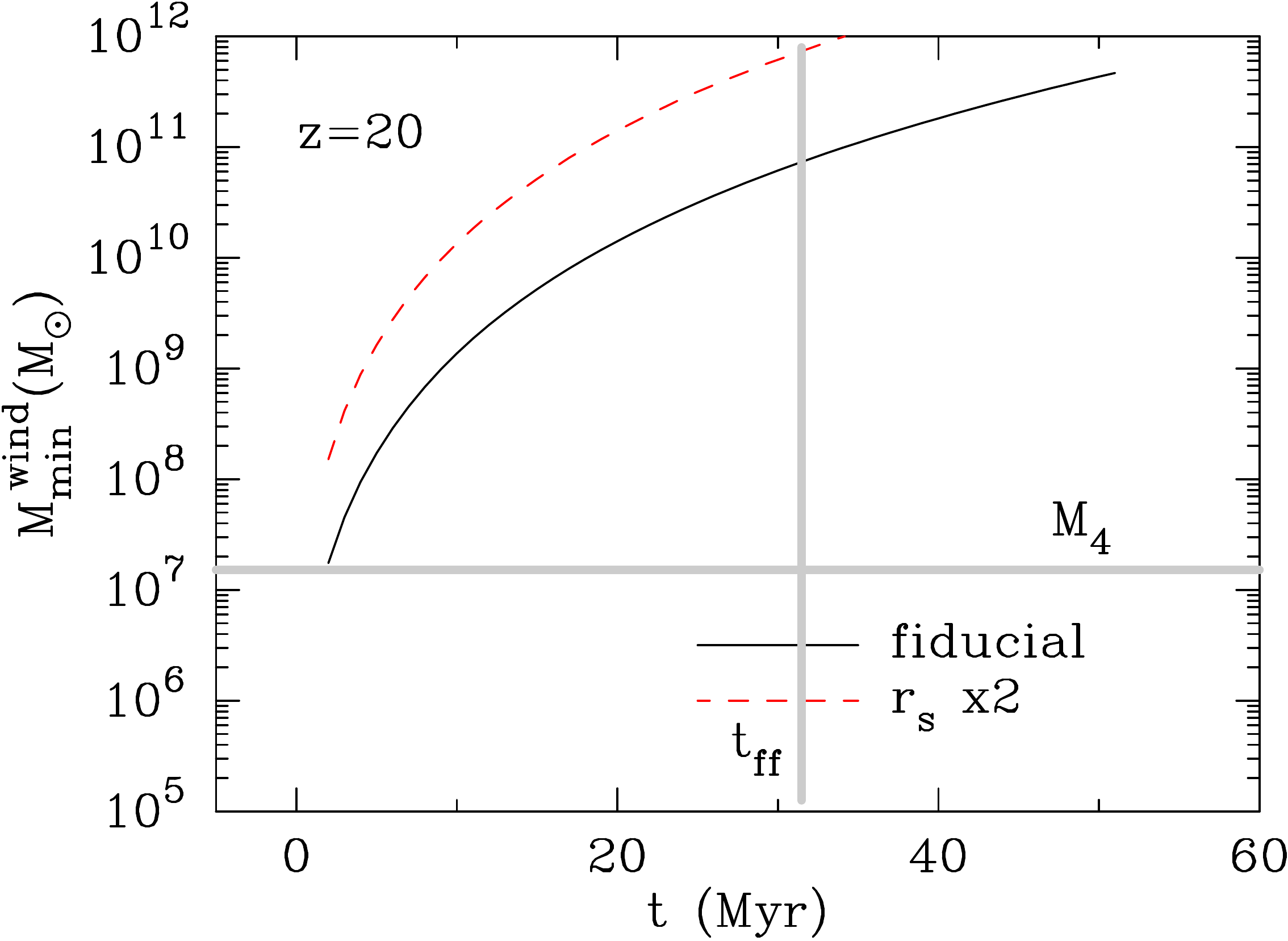,angle=270,width=8.5truecm}}}
\caption{Time-dependence of the minimum dark matter halo mass, $M^{\rm wind}_{\rm min}$, that is surrounded by a region in which DCBH formation can occur. The {\it black solid line}/{\it red dashed line} assumes that $r_{\rm s}$ is given by Eq~(\ref{eq5})/twice that. This Figure demonstrates that in our model outflows sterilize putative DCBH formation sites in dark matter halos with $M \lsim 6 \times 10^{10}\hs M_{\odot}$.} 
\label{fig:minmass}
\end{figure}

Eq~(\ref{eq5}) shows that the radius of pollution scales with stellar mass as $r_{\rm s} \propto M_*^{1/5}$ at a fixed time. Similarly, the radius at which\footnote{This assumes that $\jlw$ is dominated by a single nearby source, i.e. that there is negligible contribution from the background.} $J_{\rm LW}(r,t)=\jcrit$, denoted with $r_{\rm 300}$, scales as $r_{\rm 300} \propto M^{1/2}_*$ (Eq~\ref{eq2}, where $M_* \propto M$). If we set these two radii equal, we obtain a (time-dependent) minimum stellar -- and hence dark matter halo -- mass that can be surrounded by pristine region in which  $J_{\rm LW}(r,t)>\jcrit$. The {\it solid line} in Figure~\ref{fig:minmass} shows this minimum mass, $M^{\rm wind}_{\rm min}$, as a function of time at $z=20$. In our models, we evaluate $r_{\rm s}$ at one free-fall time (this choice is motivated by the physical argument that the collapsing gas cloud must be exposed to $\jlw >\jcrit$ during the entire time it takes to collapse into a black hole, see \S~\ref{LWLum}), $t_{\rm ff}$, which is indicated by the {\it grey vertical line}. The plot shows that at $z=20$ only dark matter halos with $M \gsim 6 \times 10^{10}\hs M_{\odot}$ are surrounded by a region in which DCBH can form. This mass is about two orders of magnitude larger than the mass associated with $T_{\rm vir}=10^4$ K (represented by the {\it grey horizontal line}).  The {\it red dashed line} shows the time dependence of the minimum mass under the assumption that $r_{\rm s}$ is twice that given by Eq~(\ref{eq5}).
\subsection{Genetic Metal Pollution}
\label{sec:genetic}

Metal enrichment of galaxies can proceed via ``genetic'' heritage \citep{Schneider06a} of metals from lower mass progenitors, rather than through outflows from neighbors. The quantity $P_{\rm gen}(M_{\rm 4}[z])$ denotes the probability that a halo {\it did not} inherit any metals from any of its progenitor halos. We take this latter probability from Trenti \& Stiavelli (2007, 2009), who used linear theory to compute the probability that a halo of mass $M$ collapsing at $z$ had a progenitor halo of a mass $M_{\rm prog} > M_{\rm H2}$ at redshift $z'>z_{\rm min}$. Here, $M_{\rm H2}$ denotes the minimum halo mass in which gas can cool via \HH-cooling (and subsequently form a star). The redshift $z_{\rm min}=z+\Delta z$, where $\Delta z$ is the change in redshift during the time that elapsed between the collapse of the progenitor halo, the formation and death of its star(s). In other words, the formation and evolution of the star(s) introduces a delay in the formation of a progenitor halo and the metal enrichment it can introduce. In detail, $P_{\rm gen}(M_{\rm 4}[z])$ depends on the LW-backgrounds and their redshift evolution, as they affect the minimum mass in which gas can cool (i.e. $M_{\rm H2}=M_{\rm H2}(z,\jbg)$, see Trenti \& Stiavelli 2009 and references therein). We take their fiducial model shown in Figure~1, which predicts that $P_{\rm gen}(M_{\rm 4}[z=10])\sim 1.0 $ and that it decreases to $P_{\rm gen}(M_{\rm 4}[z=20])\sim 0.1$. However, this choice has only a minor impact on our results, as we show that there will be much larger uncertainties associated with some of our other model parameters.

\section{LW-Background}
\label{sec:lwbg}
\subsection{Lyman-Werner Luminosity of Galaxies}
\label{LWLum}

The LW background is generated by hot, young stars in star forming galaxies. We assume an instantaneous burst of star formation taking place in a galaxy (i.e. the illuminating source) of gaseous mass $M_g = (\Omega_b/\Omega_m) M$ in a fraction $\epsilon_{\rm DC}$ (the `duty cycle') of all dark matter halos. During this burst a fraction $f_\star$ of gas is turned into stars, therefore yielding a total stellar mass $M_\star=f_\star M_g$. We constrain the value of $f_\star=0.05$ using the UV-LF of $z\sim 8$ drop-out galaxies (see Appendix~\ref{sec:Lcal}).
 
We assume that stars form according to a Salpeter IMF in the range ($m_{low}, m_{up}$) = ($1 \msun, 100 \msun$)
and with absolute metallicity $Z=10^{-3}$.  Under these assumptions\footnote{We discuss the impact of these assumptions in \S~\ref{Sum}.}, the {\it mean} LW photon production rate per solar mass of stars formed, $\langle Q_{LW} \rangle $, can be computed exactly from population synthesis models: we use here \texttt{Starburst99}\footnote{http://www.stsci.edu/science/starburst99/} by
\cite{Leitherer99}. The time dependence of the production rate of LW (integrated over the LW band $11.2-13.6$ eV) photons under these conditions is well approximated by a simple analytical form, 
\be 
\langle Q_{LW}(t) \rangle =Q_0 [(1+(t_6/4)]^{-3/2} e^{-t_6/300} 
\label{QLW}
\ee 
with $Q_0$ = $10^{47} \rm s^{-1} \msun^{-1}$ and $t = 10^6 t_6$ yr. We explicitly compare Eq~\ref{QLW} to the \texttt{Starburst99} output in Appendix~\ref{app:fit}.
Eq. (\ref{QLW}) illustrates the important point that after $\sim 4$ Myr, the production rate of LW photons rapidly drops
as a result of the death of short-lived massive stars. 

The LW intensity (in erg s$^{-1}$ cm$^{-2}$  Hz$^{-1}$ sr$^{-1}$) of such a source at a distance $r$ at time $t$ is given by:
\be
\langle J^{\rm 1s}_{\rm LW}(r,M,t_{\rm ff}) \rangle=\frac{\langle L _{\rm LW} (M, t_{\rm ff}) \rangle}{16\pi^2r^2}f_{\rm mod}(r),
\label{eq2}
\ee where the mean LW luminosity density $\langle L _{\rm LW}(M, t) \rangle$ (in erg s$^{-1}$ Hz$^{-1}$) relates to $\langle Q_{LW}(t) \rangle$
\be
\langle L _{\rm LW} (M, t) \rangle= \frac{h\langle \nu \rangle}{\Delta\nu}\langle Q_{LW}(t) \rangle f_{\rm esc, LW}\left(\frac{M_\star}{M_\odot}\right) 
\label{eq:lum}
\ee
where $\langle \nu \rangle$ is the mean frequency of the LW energy band of width $\Delta\nu$. The quantity $f_{\rm esc, LW}$ denotes the escape fraction of LW photons from the galaxy. Our fiducial calculation assumes that the escape fraction of LW photons is $f_{\rm esc,LW}=1$. This assumption may not be true, and we investigate the impact of different $f_{\rm esc,LW}$ in \S~\ref{Formation}. Finally, the term $f_{\rm mod}(r)$ describes the extra dimming introduced by the LW-horizon. We take the fitting formula for $f_{\rm mod}(r)$ from \citet{Ahn}, who derived that $f_{\rm mod}(r)=1.7\exp\big{[}-\big{(} r_{\rm cMpc}/116.29\alpha\big{)}^{0.68} \big{]}-0.7$ if $r_{\rm cMpc}/\alpha \leq 97.39$ and zero otherwise. Here,  $r_{\rm cMpc}$ is the distance to the source in comoving Mpc, and $\alpha=\Big{(}\frac{h}{0.7} \Big{)}^{-1}\Big{(}\frac{\Omega_{\rm m}}{0.27} \Big{)}^{-1/2}\Big{(}\frac{1+z}{21} \Big{)}^{-1/2}$ \citep[see][]{HF12}.

Finally, Eq~\ref{eq2} shows that we evaluate $\langle L_{\rm LW}(M,t) \rangle$ at one free-fall time\footnote{The free-fall time is $t_{\rm ff}=\sqrt{\frac{3\pi}{32\ G \rho}}$ \citep[e.g.][]{BT87}, where we assumed $\rho \sim 200 \bar{\rho}$.}, $t_{\rm ff} \sim 83([1 + z]/11)^{3/2}$ Myr, after the burst. This choice is motivated by the physical argument that the collapsing gas cloud must be exposed to $\jlw >\jcrit$ during the entire time it takes to collapse into a black hole (but see Fernandez et al. 2014 and the discussion in \S~\ref{Sum}). This time-scale corresponds approximately to the free-fall time. In this scenario, it is most optimistic (for creating DCBH sites) to assume that the burst occurred exactly when the gas in the target halo started its collapse. This assumption yields a one-to-one relation between dark matter halo mass $M$, and average LW luminosity $\langle L_{\rm LW}(M,t_{\rm ff}) \rangle$. Following D08 we assume that there exists a log-normal dispersion in $L_{\rm LW}(M,t_{\rm ff})$ around the mean $\langle L_{\rm LW}(M,t_{\rm ff}) \rangle$ with a standard deviation of $\sigma_{\rm LW}=0.4$ (which corresponds to 1 magnitude).

\subsection{Lyman-Werner Background}
\label{subsec:bg}

Once the LW-luminosity of individual galaxies has been fixed, we can compute the LW-background as \citep[e.g.][]{Ahn,HF12}
\begin{equation}
\jbg(z)=\frac{(1+z)^2}{4\pi}\int_{z}^{z+z_{\rm LW}}\frac{cdz'}{H(z')}\epsilon_{\rm LW}(z')f_{\rm mod}(z-z'),	
\label{eq:J2}
\end{equation} where the function $f_{\rm mod}(z'-z)$ takes into account the frequency dependent LW-horizon size\footnote{The relation between comoving separation and redshift is given by $dr_{\rm comoving}=cdz/H(z)$.}. We take this fitting function from \citet{Ahn}. Moreover, $\epsilon_{\rm LW}(z)$ denotes the LW-volume emissivity at redshift $z$, which is given by
\begin{equation}
\epsilon_{\rm LW}(z)=B_{\rm scat}\int_{m_{\rm min}}^{\infty} dM \frac{dn_{\rm ST}}{dM}\langle L_{\rm LW}(M,z) \rangle,
\label{eq:epsilonLW}
\end{equation} where ${dn_{\rm ST}(M, z)}/{dM}$ denotes the halo mass function which gives the number density of halos of mass $M$ (in units of comoving Mpc$^{-3}$), and $m_{\rm min} = 4 \times 10^7([1+z]/11)^{-3/2} M_{\odot}$. That is, we assume (following D08) that no star formation occurs in `minihalos' with\footnote{Following D08, this conversion assumes $\mu=1.2$, as is appropriate for fully neutral gas.} $T_{\rm vir} <10^4$ K. For our fiducial model, this translates to a maximum absolute UV-magnitude of $M_{\rm UV,max}=-10.7$ at $z=10$. We further have $B_{\rm scat}=\exp([\sigma_{\rm LW}^2 \ln^2 10]/2)\sim 1.5$ which accounts for the log-normal dispersion in $L_{\rm LW}(M,z)$.

\begin{figure}
\vbox{\centerline{\epsfig{file=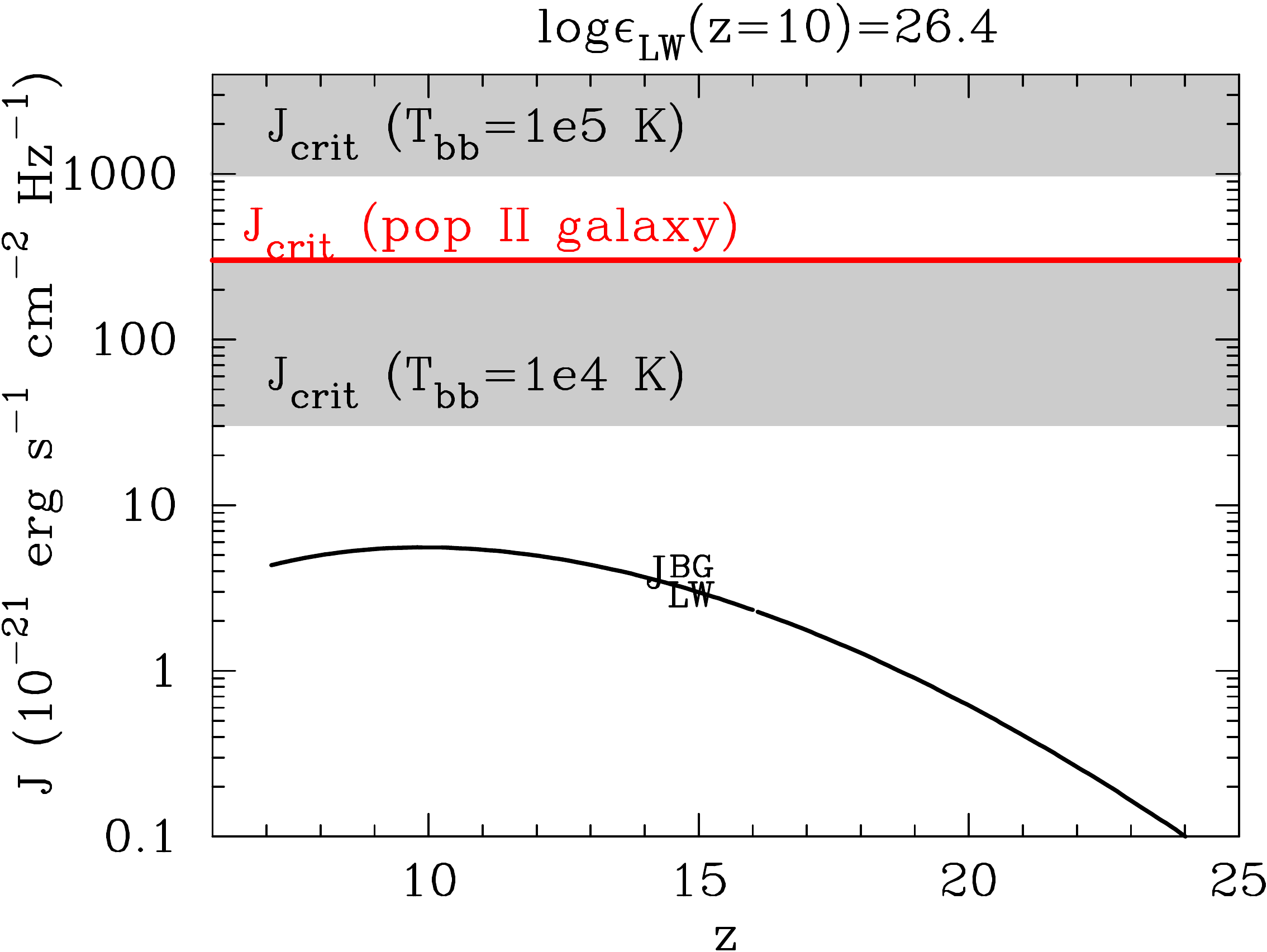,angle=270,width=8.5truecm}}}
\caption{Redshift evolution of $\jbg$ (assuming $\log \epsilon_{\rm LW}(z=10)=26.4$). For comparison, {\it shaded grey regions} show $J_{\rm crit}=30-300$ (derived for a $T=10^4$ K blackbody spectrum) and $J_{\rm crit} \geq 10^3$ (derived for a $T=10^5$ K blackbody spectrum). We have also shown our assumed $\jcrit=300$ as the {\it horizontal red line}.} 
\label{fig:bg}
\end{figure}

Figure~\ref{fig:bg} shows the redshift evolution of $\jbg(z)$. Figure~\ref{fig:bg} shows the predicted $z-$evolution of $\jbg$. The {\it grey regions} show $\jcrit$ derived for a $T=10^4$ and $T=10^5$ K black body. The {\it solid red horizontal line} shows our adopted (intermediate) $\jcrit=300$. Figure~\ref{fig:bg} shows that our predicted $\jbg \ll \jcrit$ at all redshifts. In \S~\ref{Stat} we study the fluctuations in the LW-background that a collapsing gas cloud can be exposed to. 


\subsection{Lyman Werner Background Fluctuations}
\label{Stat}

\label{sec:mc}
The LW-background is expected to be spatially very uniform, because of the large mean free path of LW-photons, $\lambda_{\rm mfp}=100$ cMpc (see \S~\ref{LWLum}). Any collapsing gas cloud in the Universe is therefore expected to see a large number of LW-emitting sources, which suppresses spatially fluctuations in the LW background. However, large departures from the background value do exist, especially in close proximity to bright LW emitting galaxies. 

To quantify these departures, we follow D08 and generate random realisations of star forming galaxies surrounding a putative DCBH formation site. We perform calculations in a coordinate system that is centered on the gas cloud of mass $M_4=4 \times 10^7 \hs M_{\odot}$ (which corresponds to a virial temperature $T_{\rm vir}=10^4$ K at $z=10$) that is possibly collapsing into a black hole. The environment of the cloud is sampled by $N_{r}=100$ concentric spherical shells spaced evenly in $\log r$ from $r=r_{\rm min}=2r_{\rm vir}$ out to a maximum radius $r_{\rm max}=66$ Mpc (proper). We denote the radius and thickness of shell number $j$ by $r_j$ and $dr_j$, respectively. Furthermore, the mass function ${dn_{\rm ST}(m,z)}/{dM}$ is sampled by $N_m=400$ mass bins that are spaced evenly in $\log m$ from $\log m_{\rm min}=5.0$ to $\log m_{\rm max}=15.0$. Mass bin number $i$ contains halos in the mass range $\log m_i\pm d\log m_i/2$. 

The average number, $N(M, r)dMdr$, of halos within the mass range $M \pm dM/2$ that populate a surrounding spherical shell of physical radius $r$ and thickness $dr$, is given by

\begin{equation}
N(M, r)dMdr = 4\pi r^2 (1 + z)^3\frac{dn_{\rm ST}}{dM}[1 + \xi] dMdr,
\label{eq:num}
\end{equation} where the factor $(1 + z)^3$ converts the number density of halos into proper Mpc$^{-3}$. The quantity $\xi \equiv \xi(M_4, M, z, r)$ denotes the two-point correlation function, which gives the excess (above random) probability of finding a halo of mass $M$ at a distance $r$ from the central halo\footnote{D08 required that $r>r_{\rm min}=r_{\rm vir}$, where
\be 
r_{\rm vir} = 0.784 \left(\frac{M}{10^8 h^{-1} M_\odot}\right)^{1/3} \Omega_M^{-1/3}\left(\frac{1+z}{10}\right)^{-1} h^{-1} {\rm kpc}.
\label{eq7}
\ee  This restriction meant to exclude spatial regions that reside within the virial radius of the illuminating halo. Our present analysis finds that halos at $r<r_{\rm min}$ lie inside the polluted radius anyway, which makes this (admittedly arbitrary) constraint irrelevant.}. In the formalism of D08, $\xi$ takes into account non-linear clustering of halos, which is important at close separations of dark matter halos. As stated previously, we assume that $m_{\rm min} = 4 \times 10^7([1+z]/11)^{-3/2} M_{\odot}$. It is worth stressing that the outflows sterilize putative DCBH formation sites in dark matter halos well above this mass-scale (see Fig~\ref{fig:minmass}). Apart from raising the global background, lower mass halos would not contribute to the abundance of potential DCBH formation sites in the presence of outflows. 

We use a Monte Carlo procedure to generate a large number of random realizations of the spatial distribution of dark matter halos surrounding the gas cloud of interest. For each realization, we assign LW luminosities to the field dark matter halos using Eq~\ref{eq:lum}, and compute the total LW flux illuminating the target halo from 
\begin{eqnarray}
J_{\rm LW}=\sum_{i=1}^{N_{\rm m}}\sum_{j=1}^{N_{\rm r}} \sum_{k=1}^{N(m_i,r_j)}\langle J^{\rm 1s}_{\rm LW}(r_j,M_i,t_{\rm ff})\rangle \Theta[r_j-r_{\rm s}(m_i)]\times \\ \nonumber
\times B_{\rm LW}(R_{\rm k,1}) \Theta[R_{\rm k,2}-(1-\epsilon_{\rm DC})],\label{eq:mc}
\end{eqnarray} where $\Theta(x)$ is the Heaviside step function [$\Theta(x)=0$ for $x \leq 0$, $\Theta(x)=1$ for $x>0$]. The Heaviside function in the first line ensures that we do not include halos which lie inside the radius of pollution by the neighboring halo/galaxy (i.e. $r>r_{\rm s}$, where $r_{\rm s}$ is given by Eq~\ref{eq4}). The sum over $k$ reflects the fact that any given mass-radius bin may contain $N(m_i,r_j)\equiv 4\pi r^2_jdr_j\frac{dn_{\rm ST}}{dM}dm_i>1$ halos.  The Heaviside function in the second line accounts for the finite `duty cycle' of the dark matter halos in our models.  The factor $B_{\rm LW}(R_{\rm k,1})\equiv \hs{\rm dex}(\sigma_{\rm LW}\sqrt{2}{\rm erf}^{-1}[2R_{\rm k,1}-1])$ accounts for the lognormal fluctuations in the LW-luminosity of single dark matter halos, where the $R_{k}$ variables denote random numbers between 0 and 1 (see D08).

Equation~(\ref{eq:mc}) gives the Lyman-Werner flux that is seen by a collapsing cloud in a single Monte-Carlo realization. We repeat the Monte-Carlo calculation $N_{\rm mc}$ times in order to derive an accurate PDF of $J_{\rm LW}$. We are specifically interested in the probability
\begin{equation}
P(J_{\rm LW} > \jcrit) = \frac{N_{\rm mc}(J_{\rm LW} > \jcrit)}{N_{\rm mc}},
\label{word}
\end{equation} where $N_{\rm mc}(J_{\rm LW} > \jcrit)$ denotes the number of Monte-Carlo simulations in which we found $\jlw>\jcrit$. In cases we need to compute the low-probability tail of the LW-flux PDF, we need many (potentially as large as $N_{\rm mc} \gg 10^7$) Monte-Carlo runs to sample the full PDF. Fortunately, the high-flux tail of the PDF is dominated by single nearby sources, and can be computed analytically (see D08). We discuss these analytic calculations in the Appendix.

\section{DCBH formation probability}
\label{Formation}
\begin{table*}
\centering.
\caption{Models.} 
\begin{tabular}{l l l l}
\hline\hline\\
symbol & model \# & short description & model description\\ 
\hline \\
$\triangle$ & (i) & fiducial & $\jcrit=300$; Starformation occurs in all dark matter halos with $T_{\rm vir}>10^4$ K\\ & & & ($M_{\rm UV,max} = -10.7$ at $z = 10$); $f_{\rm esc,LW}=1.0$; galactic outflows described by Eq~\ref{eq5}.\\
$\textcolor{red}{\mathlarger{\mathlarger{\mathlarger{\mathlarger{\bullet}}}}}$ & (ii)& $M_{\rm UV,max}=-14$& Same as (i), but extrapolate the UV-luminosity function to $M_{\rm UV,max}=-14.0$\\
$\textcolor{blue}{\blacksquare}$ & (iii)& $f_{\rm esc,LW}=0.5$& Same as (i), but with $f_{\rm esc,LW}=0.2$. \\
$\textcolor{green}{\mathlarger{\mathlarger{\mathlarger{\star}}}}$ & (iv) & no winds & Same as (i), but ignore galactic winds (i.e. $r_{\rm s}=0$).  \\
$\bigcirc$ & (v) & $\jcrit=100$& Same as (i), but decrease $\jcrit$ from $\jcrit=300$ to $\jcrit=100$\\
\textcolor{red}{$\bigodot$}& (vi) & $\jcrit=30$ & Same as (i) but use $\jcrit=30$ (see text for details) \\
\hline\hline
\end{tabular}
\label{table:symbolsmodels}
\end{table*}

\begin{figure*}
\vbox{\centerline{\epsfig{file=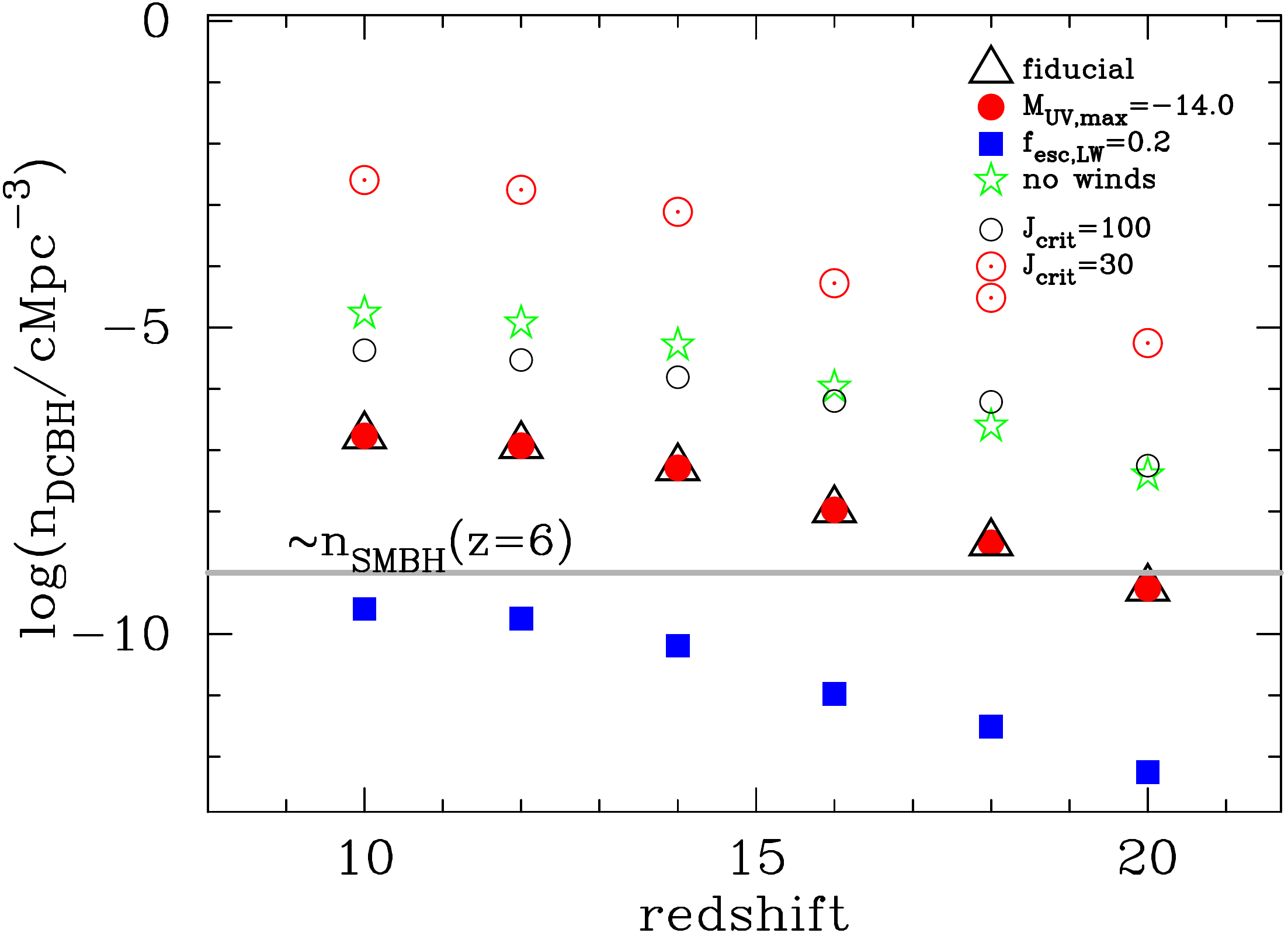,angle=270,width=13.5truecm}}}
\caption{Predicted redshift evolution of the comoving number density of putative DCBH formation sites, $n_{\rm DCBH}(z)$, for six different models, each represented by one type of data-point (see text \& Table~\ref{table:symbolsmodels} for details). For most models $n_{\rm DCBH}$ gently decreases with redshift. This Figure shows that the predicted $n_{\rm DCBH}(z)$ at a given $z$ depend on especially strongly on ({\it i}) the escape fraction of LW photons, ({\it ii}) $J_{\rm crit}$, and ({\it iii}) the presence of winds.} 
\label{fig:ncan}
\end{figure*}

We compute the number density of putative DCBH host halos, $n_{\rm DCBH}(z)$, as
\begin{equation}
n_{\rm DCBH}(z)=\int_{M_{\rm min}}^{M_{\rm max}}\frac{dn_{\rm ST}(z)}{dM}P(\jlw>\jcrit|z)P_{\rm gen}(M,z)dM,
\end{equation} where $P(\jlw>\jcrit|z)$ is given by Eq~\ref{word}, and $P_{\rm gen}(M,z)$ is described in \S~\ref{sec:genetic}.
We approximate this expression as
\begin{equation}
n_{\rm DCBH}(z) \approx n(M>M_{\rm 4}[z])P(\jlw>\jcrit|z)P_{\rm gen}(M_{\rm 4}[z],z).
\end{equation} 
Here, $M_{\rm 4}(z)$ denotes the halo mass at which $T_{\rm vir}=10^4$ K. We integrate over the (redshift dependent) dark matter halo mass function to compute $n(M>M_{\rm 4}[z])$. We point out that the redshift dependence of $P(\jlw>\jcrit|z)$ is embedded in the halo mass function evolution, the mass-dependent radius of the enriched bubbles [Eq.~\ref{eq5}; we evaluate this expression at $t_{\rm ff}(z)$], the relation between halo mass and LW luminosity [Eq.~\ref{eq2}; evaluated at $t_{\rm ff}(z)$], and the two-point correlation function\footnote{We have approximated the redshift dependence of $\xi$ by simply rescaling the two-point function computed at $z=10$ with the linear growth factor $D_+(z)$ as 
\begin{equation}
\xi(z) = \xi(z=10)\left[\frac{D_+(z)}{D_+(z=10)}\right]^2.
\end{equation}.} $\xi(M_4,M,r,z)$. \\

Figure~\ref{fig:ncan} shows the predicted $n_{\rm DCBH}(z)$ at $z=10,12,14,...$ for six different models, which are described in Table~\ref{table:symbolsmodels}. In each model, we change one parameter (i.e. the maximum $M_{\rm UV,max}$ to which we extrapolate the UV luminosity function, $f_{\rm esc,LW}$, $\jcrit$, and $r_{\rm s}$).  
The main purpose of model (vi), the model with $\jcrit=30$, is to facilitate comparison with recent works by \citet{Agarwal12,Ag14}. For each model, we compute $P(J_{\rm LW} > \jcrit)$ using both Monte-Carlo calculations (see \S~\ref{sec:mc}) and analytic calculations (Appendix~\ref{sec:fluxpdf}). When $P(J_{\rm LW} > \jcrit)$ is very small (i.e. when $P(J_{\rm LW} > \jcrit) \ll 10^{-6}$), we only use the analytic solutions. Our results are shown in Figure~\ref{fig:ncan}. We will not discuss each model. Instead, we list the most important points that can be taken away from this analysis. These include:

\begin{itemize}

\item At each redshift, our models generally predict a range of number density of putative DCBH formation sites, which span $\sim$ 5 orders of magnitude depending on $f_{\rm esc,LW}$, $\jcrit$, and whether winds are included or not. We find that  $n_{\rm DCBH}(z=10)$ decreases with redshift from $n_{\rm DCBH}(z=10) \sim 10^{-10}-10^{-5}$ cMpc$^{-3}$ to $n_{\rm DCBH}(z=20) \sim 10^{-12}-10^{-7}$ cMpc$^{-3}$, which should be compared to the observed number density of SMBHs at $z=6$, $n_{\rm SMBH}(z\sim6) \sim10^{-9}$ cMpc$^{-3}$. The model with $\jcrit=30$ typically predicts $n_{\rm DCBH}(z)$ to lie orders of magnitude higher (we discuss this in \S~\ref{Sum} below). 

\item Our predicted $n_{\rm DCBH}(z)$ does not depend on where we truncate the UV luminosity function. This somewhat peculiar result is a consequence of having $\jcrit=300$, which requires a nearby LW-luminous galaxy. For example, the {\it left panel} of Fig~\ref{fig:pdf2} shows that at $z=10$ we require a neighbour at a few kpc with $M_{\rm UV} \lsim -18$, even in the absence of winds. We therefore do not care about the number density of fainter (lower mass) galaxies. Galactic winds eliminate preferentially close pairs of halos as putative DCBH sites (at $r \lsim 20-30$ kpc, see the {\it left panel of Fig~\ref{fig:pdf2}}). This increases the intrinsic LW luminosity of a galaxy that is required to reach $\jcrit$ even further to $M_{\rm UV} \sim -21 \pm 1$ (see the {\it right panel} of Fig~\ref{fig:pdf2} in the Appendix).

\item Winds reduce the predicted $n_{\rm DCBH}$ by about 2 orders of magnitude, which suggests that metal enrichment by winds is more important than genetic enrichment (which affect the predicted  $n_{\rm DCBH}$ at less than 1 order of magnitude, see \S~\ref{sec:genetic}). The precise impact depends on the wind implementation and also on $J_{\rm crit}$. This dependence is best illustrated in Fig~\ref{fig:pdf1} in the Appendix, which shows that ({\it i}) increasing $r_{\rm s}$ by a factor of $\sim 2$ reduces $P(\jlw > \jcrit)$ by an additional $\sim1.5$ orders of magnitude, and ({\it ii}) winds affect the especially the high-$\jlw$ end of the $\jlw$-PDF.
We find that winds remove the closest pairs of halos as possible DCBH formation sites, but allow for DCBH formation around the most UV-luminous galaxies, that are hosted by the most massive halos, i.e. with $M>M_{\rm min}^{\rm met}(z)$ (see Fig~\ref{fig:minmass}, and also Fig~\ref{fig:pdf2} in the Appendix). 

\item The predicted $n_{\rm DCBH}(z)$ increases with cosmic time for all models, but saturates at $z \lsim 12$. This due to ({\it i}) the fact that $P_{\rm gen}(M_{\rm 4}[z])$ reaches unity at $z \lsim 12$, and ({\it ii}) the competition between the $z-$evolution in the halo mass function, in which the number density of dark matter halos of fixed mass $M$ decreases with $z$, and the LW luminosity-to-mass ratio which increases towards higher redshift. Because of this competition, the probability to have a nearby neighbour with some luminosity $L_{\rm LW}$ evolves only weakly with $z$. 
\end{itemize}

We stress that we have further confirmed that our results barely depend on the assumed duty cycle, $\epsilon_{\rm DC}$, and weakly on the assumed $\sigma_{\rm LW}$, provided that we fit these models to the observed UV-luminosity functions (also see D08).

\section{Summary and discussion}
\label{Sum}
In this paper we have summarized the properties - and computed the abundance - of halos in which favorable conditions for the formation of `Direct Collapse Black Holes' (DCBHs) are present. DCBH formation involves the collapse of a gas cloud directly into a very massive, $m_0 \approx 10^{4-6} \Msun$, black hole. DCBH formation poses stringent requirements on the putative host halos: ({\it i}) the halo virial temperature has to be $>10^4$ K to ensure the high accretion rates allowed by atomic cooling; ({\it ii}) the gas metallicity must be $Z < Z_{crit} =10^{-5\pm 1} Z_\odot$ to prevent fragmentation into clumps induced by heavy elements and dust cooling; ({\it iii}) they must be exposed to a UV intensity $\jlw  > \jcrit =300$ to strongly depress the abundance of \HH molecules, also acting as cooling agents. Assessing when and where these conditions are met through cosmic evolution has been the primary goal of this study. \\

We stress that our adopted $\jcrit=300$ is intermediate between $\jcrit=30-300$ (appropriate for a $T=10^4$ K blackbody spectrum, see Shang et al. 2010) and $\jcrit=10^3$ (appropriate for a $T=10^5$ K blackbody spectrum, see Wolcott-Green et al. 2011). In \S~\ref{subsec:bg} we showed that the LW-background value was likely $\jbg \ll \jcrit$.
We also studied fluctuations in the LW-background, and computed PDFs of $J_{\rm LW}$ impinging on a collapsing gas cloud inside a dark matter halo with $T_{\rm vir}=10^4$ K. Previous works had demonstrated that $J_{\rm LW}$ can be elevated significantly in close proximity to star forming galaxies. In this work we introduced physically motivated prescriptions to estimate metal pollution in these regions via galactic outflows (based on the models by Madau et al. 2001, see \S~\ref{Metal}). We found that metal pollution strongly affects the high  $J_{\rm LW}$-tail of the-PDF as it `sterilizes' putative DCBH host halos at close separations $\simlt 10$ kpc. This preferential removal of close-halo pairs as putative DCBH formation sites requires the illuminating LW-source to be intrinsically bright in LW and hence the UV continuum. Our fiducial model predicts DCBH formation to occur around galaxies with $M_{\rm UV}\sim -21 \pm 1$ (see Appendix~\ref{appendix}). Interestingly, this suggests that DCBH formation is sensitive to the bright end of the observed UV-LF.\\

We concluded with an estimate of the number density of putative DCBH formation sites, $n_{\rm DCBH}(z)$. Our analysis thus includes ({\it i}) fluctuations in the LW-background, taking into account the reduction of the high-end tail in the $\jlw$-PDF caused by galactic winds, and ({\it ii}) the `genetic' enrichment probability, which is the probability that the halo that is collapsing has been enriched by star formation that occurred inside a progenitor halo (see \S~\ref{sec:genetic}). We find that $n_{\rm DCBH}$ increases with cosmic time from $n_{\rm DCBH}(z=20) \sim 10^{-12}-10^{-7}$ cMpc$^{-3}$ to $n_{\rm DCBH}(z=10) \sim 10^{-10}-10^{-5}$ cMpc$^{-3}$. Galactic winds suppress the predicted $n_{\rm DCBH}(z)$ by $\sim 2-2.5$ orders to magnitude. Metal enrichment by winds therefore dominates over `genetic' enrichment, which reduces $n_{\rm DCBH}(z)$ by less than 1 order of magnitude. For comparison, the observed number density of SMBHs at $z=6$, $n_{\rm SMBH}(z\sim6) \sim10^{-9}$ cMpc$^{-3}$. Since we do not expect each DCBH to grow into a SMBH, we likely require $n_{\rm DCBH}(z) \gg n_{\rm SMBH}$ in order for DCBH formation to provide a viable mechanism for SMBHs. Our analysis clearly highlights the dependence of our predictions on ({\it i}) the escape fraction of LW photons, ({\it ii}) $\jcrit$, and ({\it iii}) the galactic outflow prescription.\\

We note that our results differ from the predictions by Agarwal et al. (2012, 2013, 2014) mostly because of their choice $\jcrit=30$: Agarwal et al. (2012) predict $n_{\rm DCBH}(z=10)\sim 3\times 10^{-3}$ cMpc$^{-3}$ for $f_{\rm esc,LW}=1.0$. This corresponds exactly to predictions of our model (vi) that also adopts $\jcrit=30$, which is remarkable given the vastly different approaches that we used\footnote{Agarwal et al. (2012) do not include galactic winds in their analysis, in contrast to our model (vi). However, $P(\jlw > \jcrit)$ only decreases by $\sim 0.5$ orders of magnitude for $\jcrit=30$ in our models, which suggests that winds do not affect the agreement. Agarwal et al. (2014) do include winds, but they do not provide predicted number densities. However, the fact that putative DCBH formation sites exist in their 64 cMpc$^{3}$ simulation suggests that the number density is not much affected by galactic winds. This is again consistent with our analysis, which suggests that galactic winds are more important for larger $\jcrit$.}.\\

There are several caveats and additional questions raised by our study that will need additional scrutiny.
\begin{itemize}

\item The LW photon escape fraction from halos plays a key role in determining $n_{\rm DCBH}(z)$. \citet{Kitayama04} studied the mass dependence of $f_{\rm esc,LW}$ by using a simplified model for the LW absorption, and found that it increase from $f_{\rm esc,LW}=0$ to $f_{\rm esc,LW}=1$ above some critical mass \citep[this resembles the mass-dependence of the escape fraction of ionising photons as in][]{Ferrara12}.  A (strong) mass-dependence of $f_{\rm esc,LW}$ can have a major impact on our results. Surprisingly, there exists very little additional calculations of $f_{\rm esc,LW}(M)$ in the literature.

\item The precise value of $\jcrit$ is also uncertain. Here we have assumed $\jcrit=300$, intermediate between $\jcrit=30-300$ (appropriate for a $T=10^4$ K blackbody spectrum)\footnote{After this paper was submitted, \citet{Latif14} posted a preprint in which they found $\jcrit=400-700$ for a $T=10^4$ K blackbody spectrum. This underlines that $\jcrit$ is uncertain, even for a fixed spectrum.} and $\jcrit=10^3$ (appropriate for a $T=10^5$ K blackbody spectrum). The proper value of $\jcrit$ for the spectrum of a galaxy that contains population II stars is not published yet and is thus still uncertain. Furthermore, \citet{VS13} have shown that $\jcrit$ can be reduced by a factor of $\sim 10$ in the presence of a strong magnetic field and/or strong turbulence.

\item Additional uncertainties come from the simplified outflow treatment, which are assumed to be ({\it i}) spherically  symmetric, ({\it ii}) adiabatically  evolving, ({\it iii}) propagating into a medium of mean constant overdensity $\Delta = 60$. These assumptions might be overcome by a numerical study. However, we do expect our results to be only very mildly affected by a better treatment as metal pollution has been show to suppress DCBH formation only in rare regions of extremely high LW flux (see Fig. \ref{fig:pdf1}).  In addition, if massive $>100 \msun$ PopIII stars were present at those epochs, they should collapse into stellar black hole essentially swallowing all nucleosynthetic products of the progenitor star. 

\item Our analysis used \texttt{Starburst99} to generate spectra for galaxies, and assumed a metallicity of $Z=10^{-3} Z_{\odot}$. For metal-free gas, the total boost can be an order of magnitude \citep[e.g.][]{Ciardi05,Trenti09}. Changing the metallicity of galaxies thus changes their LW-luminosity. Uncertainties this introduces in the LW-volume emissivity, $\epsilon_{\rm LW}(z)$, propagate into uncertainties in $\jbg$, and thus $n_{\rm DCBH}(z)$. However, these changes are subject to the observational constraints at $z=8-10$. For example, decreasing the metallicity of {\it all} galaxies at $z=10$ would be compensated for by decreasing $f_*$, which would preserve our results (also see Appendix~\ref{sec:Lcal}). 

\item We assume a dispersion of 1 magnitude in the relation between $L_{\rm LW}$ and halo mass $M$ (see \S~\ref{LWLum}, which corresponds to a lognormal dispersion in $L_{\rm LW}(M)$ with $\sigma_{\rm LW}=0.4$). This dispersion can reflect a dispersion in metallicity, but also a dispersion in the time since the starburst that occurred in the nearby halo. We evaluate $L_{\rm LW}$ of galaxies one free-fall time after the starburst (as we motivated in \S~\ref{LWLum}). If we were to evaluate $L_{\rm LW}$ at a time $t \ll t_{\rm ff}(z)$, then the collapsing gas cloud can be exposed to $\jlw < \jcrit$ during other (earlier or later) stages of the collapse, and molecules might form which can cool the gas and induce fragmentation. Recently however, Fernandez et al. (2014) have shown that DCBH formation can occur\footnote{This is different than UV-free collapse discussed in \S~\ref{Mot}: in this scenario we require the collapsing gas cloud to be illuminated by $\jlw > \jcrit$ for at the initial stages o the collapse.} when $\jlw$ drops below $\jcrit$ during the isothermal collapse of the cloud, provided the density has reached a `point-of-no-return' value of $n \sim 10^4$ cm$^{-3}$ (beyond which collisional dissociation of H$_2$ keeps its abundance low, also see Inayoshi \& Omukai 2012). This implies that for a fraction of galaxies it would have been allowed to evaluate $L_{\rm LW}(M)$ at a somewhat earlier time, namely the time it took for the collapsing pristine cloud to reach the point-of-no-return. This can increase the dispersion in the $L_{\rm LW}-M$ relation, which can boost $n_{\rm DCBH}$ somewhat (see D08 for plots showing the impact of varying $\sigma_{\rm LW}$).

\item A final question remains also on the ability of outflow to pollute a neighboring DCBH halo host, although this mechanism is very often advocated in the literature.  Mixing the heavy elements transported by the outflow impinging onto a target halo is a very complex physical mechanism \citep{Ferrara12, Cen08},  whose solution requires sophisticated numerical schemes. \citet{Cen08} pointed out that the highly over-dense halo gas may be more robust and thus resistant to mixing with metals carried by intergalactic shocks. 
\end{itemize}

Hence, a final answer requires to solve the problem in a fully self-consistent manner, a task that is beyond present-day numerical simulation capabilities. Alternatively, it is interesting to predict observational signatures of galaxies hosting DCBHs, and to investigate whether observations can constrain this uncertain, but interesting, astrophysics \citep[see e.g.][]{DW06,Agarwal13}.

\section*{Acknowledgments} 
Support from the SNS Visitor Program (MD) is kindly acknowledged. We thank Zoltan Haiman \& Eli Visbal for helpful discussions, and an anonymous referee for a constructive report which improved the content of this paper.


\appendix

\section{Fitting Formula}
\label{app:fit}

\begin{figure}
\includegraphics[width=9.0cm,angle=0]{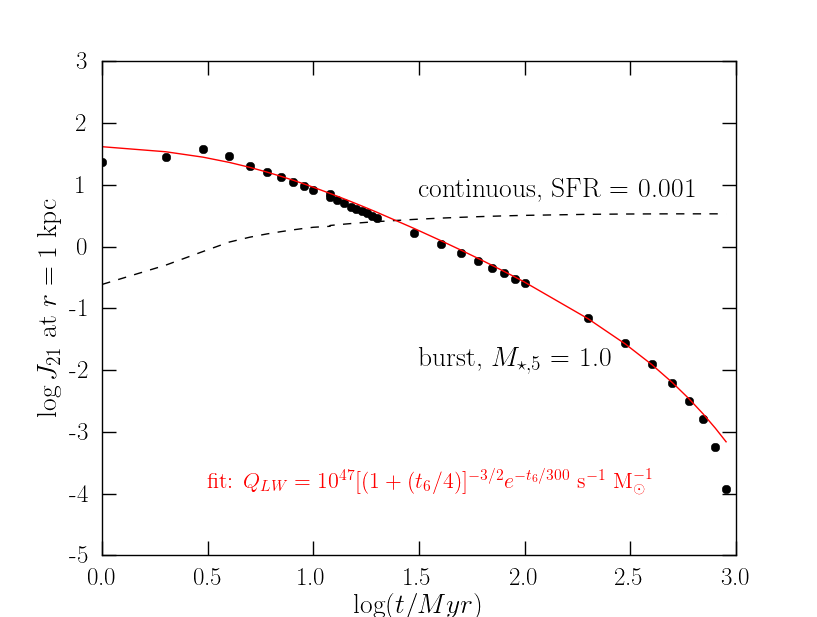}
\caption{Comparison between our fitting formula (Eq~\ref{QLW}, {\it red solid line}) and the output from \texttt{Starburst99} ({\it black filled circles}).} 
\label{fig:fit}
\end{figure}
We adopted a fitting formula in Eq~\ref{QLW} for the time-dependence of the rate at which LW-photons are emitted by a galaxy undergoing a starburst at $t=0$.
Figure~\ref{fig:fit} explicitly compares this formula ({\it red solid line}) to the output from \texttt{Starburst99} ({\it filled black circles}) for a Salpeter IMF in the range ($m_{low}, m_{up}$) = ($1 \msun, 100 \msun$) and with absolute metallicity $Z=10^{-3}$.  Note that the vertical axis here does not contain the number of LW photons, but instead the LW flux density at $r=1$ kpc. We applied the same rescaling to both the fitting formula and the \texttt{Starburst99} output, which therefore does not affect their comparison.

\section{Luminosity Calibration}
\label{sec:Lcal}
The predicted LW flux from a halo contains parameters that are poorly known from first principles and only
weakly constrained by the data (as for example  $f_\star$). It is therefore necessary to calibrate models on
additional observables as the Luminosity Function (LF) of high-$z$ drop-out galaxies which are now obtained
with increasing precision. 

The UV LFs are usually obtained by measuring the flux in the restframe spectral region ($\lambda\approx 1400-1600$ A), whereas
the LW band extends in $\lambda=912-1107$ A. Thus we need to convert the (specific) LW into a UV luminosity, for which we assume that the UV continuum slope is $\beta=-2$ (where $L_{\lambda} \propto \lambda^{\beta}$), which is consistent with the UV colors measured of high redshift drop-out galaxies \citep[e.g.][]{FinkelsteinUV}; this yields $L_{\nu,{\rm LW}}=L_{\nu,{\rm UV}}$. 

We evaluate the LW, and therefore the UV, specific luminosity of a galaxy at one free-fall time, $t_{\rm ff} \sim 83([1+z]/11)^{-3/2}$ Myr, after the burst. This choice is motivated by the physical argument that the collapsing gas cloud must be exposed to $\jlw > \jcrit$ during the entire time it takes to collapse into a black hole. This time-scale corresponds approximately to the free-fall time. In this scenario, it is most optimistic (for creating DCBH sites) to assume that the burst occurred exactly when the gas in the target halo started its collapse. This assumption yields a one-to-one relation between dark matter halo mass $M$, and UV luminosity $L_{\rm UV}$. We can predict the models UV LF - measured as the comoving number density of star forming galaxies per unit of absolute magnitude - from

\begin{equation}
\frac{dn}{dM_{\rm UV}}=\epsilon_{\rm DC}\int_{m_{\rm min}}^{\infty}\frac{dn_{\rm ST}}{dM}dM\frac{dP(M)}{dM_{\rm UV}}
\end{equation} where ${dn_{\rm ST}(M, z)}/{dM}$ is the \citet{PS74} mass function \citep[with the modiÞcation of][and is hence labelled the `Sheth-Tormen' mass function]{ST01}, which gives the number density of halos of mass $M$ (in units of comoving Mpc$^{-3}$). Furthermore, $\frac{dP(M)}{dM_{\rm UV}}=-2.5\frac{dP(M)}{d\log L_{\rm UV}}$, in which $\frac{dP(M)}{d\log L_{\rm UV}}$ is given by the lognormal distribution with mean $\log \langle L_{\rm LW}(M,t_{\rm ff}) \rangle$ and standard deviation $\sigma_{\rm LW}=0.4$ (see D08 for more details).

\begin{figure}
\vbox{\centerline{\epsfig{file=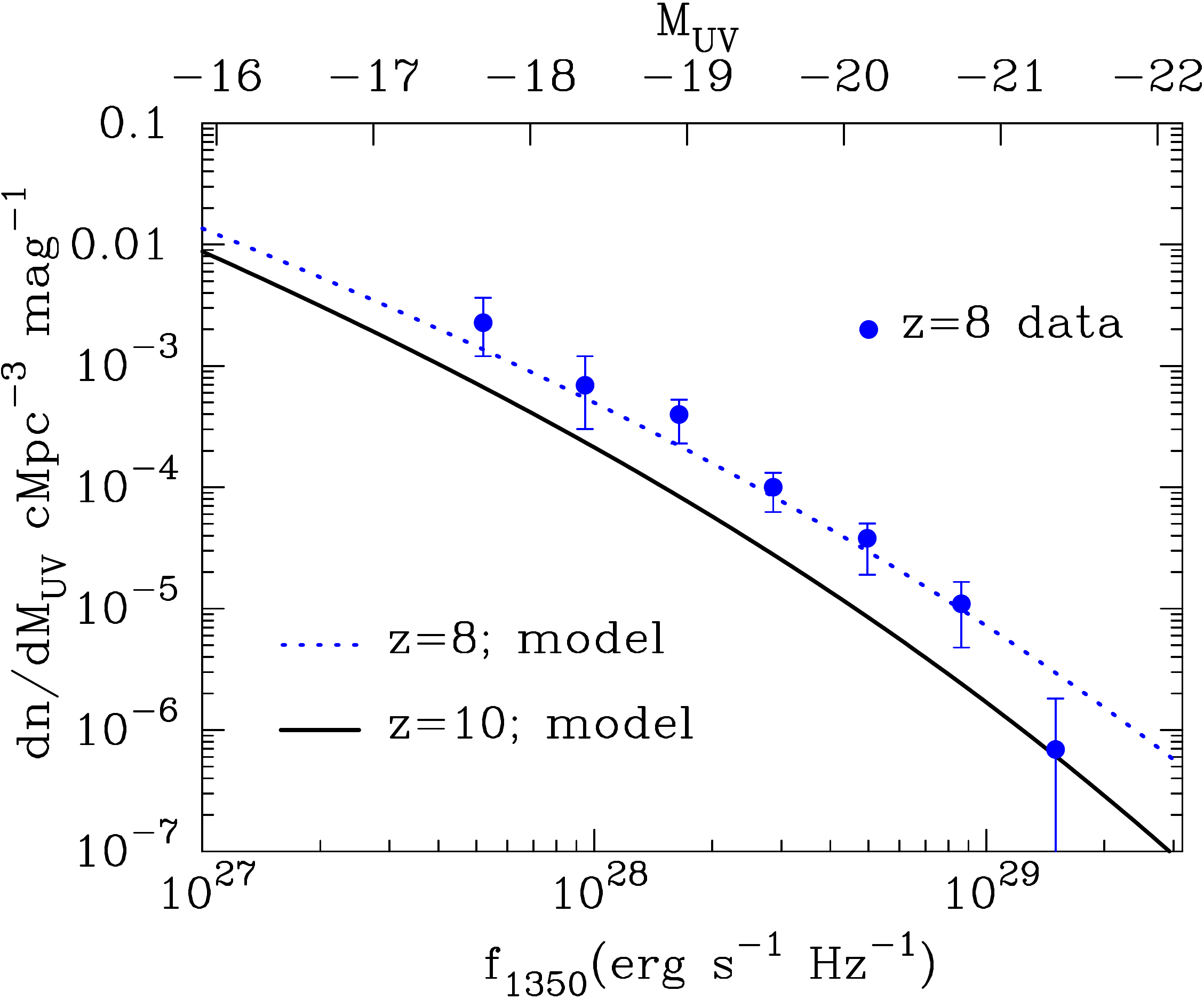,angle=270,width=8.5truecm}}}
\caption{Observed rest frame UV luminosity function
  \citep[][]{Oeschz8}  of $z\sim 8$ drop-out galaxies ({\it blue
    filled circles}) compared to that predicted by our model at $z=8$
  ({\it blue dashed line}) and $z=10$ {\it black solid line}. The
  integrated UV-luminosity density for galaxies with $L>L^{z=3}_*$ of our model at $z=10$ is $\log
  \dot{\rho}_{\rm UV}=24.5$ for $f_\star=0.05$ and $\epsilon_{\rm DC}=0.2$, consistent with observational constraints presented by \citet{Oesch12} and \citet{Bouwens13}.} 
\label{LF}
\end{figure}

Figure~\ref{LF} compares the UV luminosity function predicted by our
model at $z=10$ and $z=8$. Also shown is a recent determination of the
$z\sim 8$ drop-out luminosity function \citep[taken
from][]{Oeschz8}. Our model is clearly consistent with observational
constraints at this redshift. The $z=10$ UV luminosity density of star
forming galaxies for $f_\star=0.05$ with $L_{\rm UV}>0.06L^{z=3}_*$ in our model is $\dot{\rho}_{\rm UV}=10^{24.5}$ erg s$^{-1}$ cm$^{-2}$ Hz$^{-1}$ Mpc$^{-3}$, which lies within  $1\sigma$ of recent observational constraints by \citet{Oesch12} and \citet{Bouwens13}. We emphasis that our final results depend only little on how we assign UV luminosities to halos, provided that we extrapolate our models to the same minimum UV-luminosity, and that we constrain our models with observations. For example, if we were to evaluate the LW luminosity density of a dark matter halo at $t=0.1t_{\rm ff}$, we increase $Q_{\rm LW}$ dramatically. However, for the same value of $f_*=0.05$ we would overproduce the UV-LF at $z=8$ significantly, and we would need to reduce $f_*$. If one properly reduces $f_*$, then we get back to our original results\footnote{This is probably also the reason that our computed $\jlw$-PDF is close to that of D08. Their model was also constrained by observations.}.

\section{Analytic Calculations}
\label{appendix}

\begin{figure*}
\vbox{\centerline{\epsfig{file=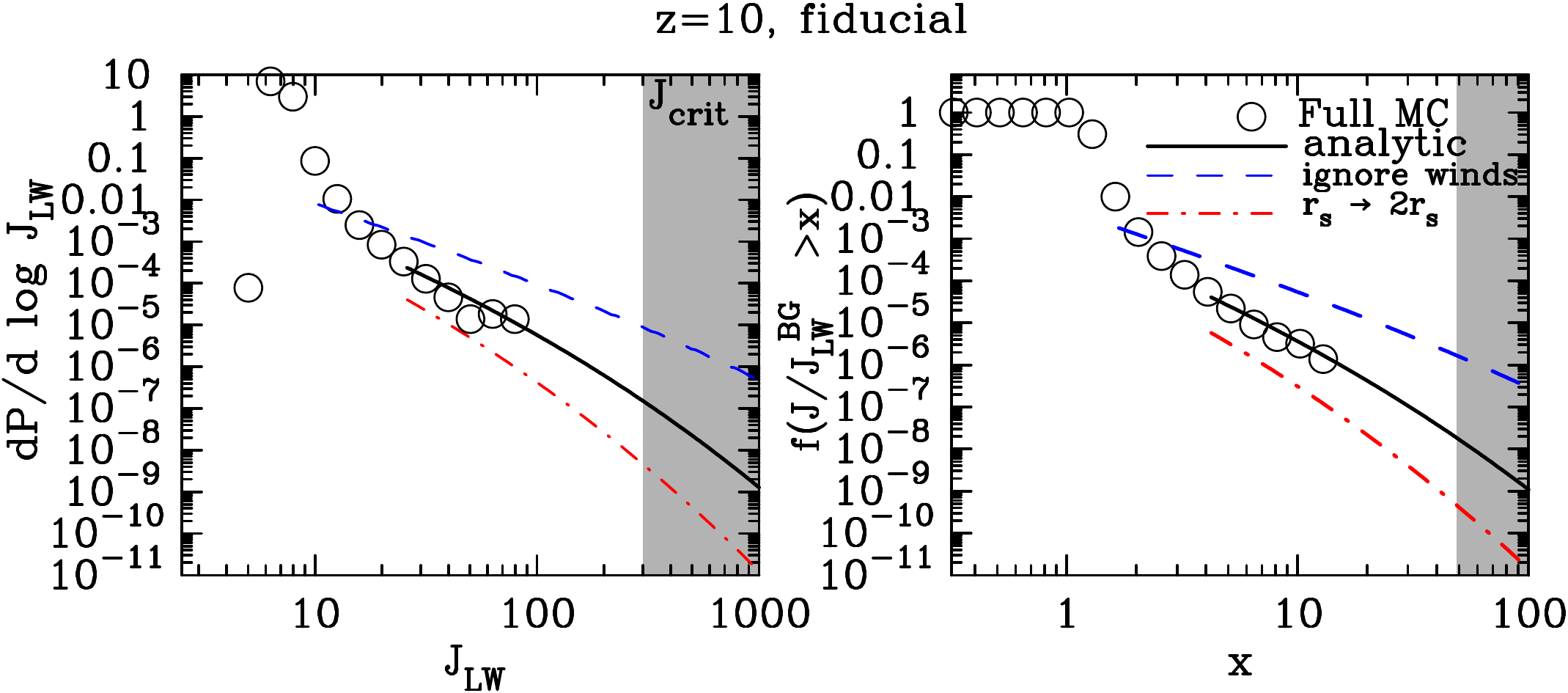,angle=270,width=14.5truecm}}}
\caption{{\it Left panel:} PDF of LW flux (measured in units of $10^{-21}$ erg s$^{-1}$ Hz$^{-1}$ cm$^{-2}$ sr$^{-1}$) permeating a collapsing halo with a virial temperature of $T_{\rm vir}=10^4$ K for the fiducial model at $z=10$. The {\it open circles} show the results from our Monte-Carlo runs. The {\it solid black line} shows our analytic calculation (Eq.~\ref{eq:jpdf}), which assumes that the total LW flux is dominated by a single nearby source. The analytic solution provides a good match to the full calculation. The {\it blue dashed line}/[{\it red dot-dashed line}] represents a model in which we ignore winds/[in which $r_{\rm s}$ is increased by a factor of $2$]. This comparison shows that the metals strongly affect the high-$\jlw$ tail of the distribution.  The {\it right panel} shows the cumulative fraction of collapsing clouds that see a boost $x$ in their LW flux compared to the background value $\jbg$. For example, this plots shows that in our fiducial model only $\sim 10^{-8}$  of all halos `see' a $\jlw$ that is $\sim 50$ times $\jbg$, which corresponds to the minimum boost required to reach $\jcrit$ (which is represented by the {\it grey regions}). } 
\label{fig:pdf1}
\end{figure*}
\begin{figure*}
\vbox{\centerline{\epsfig{file=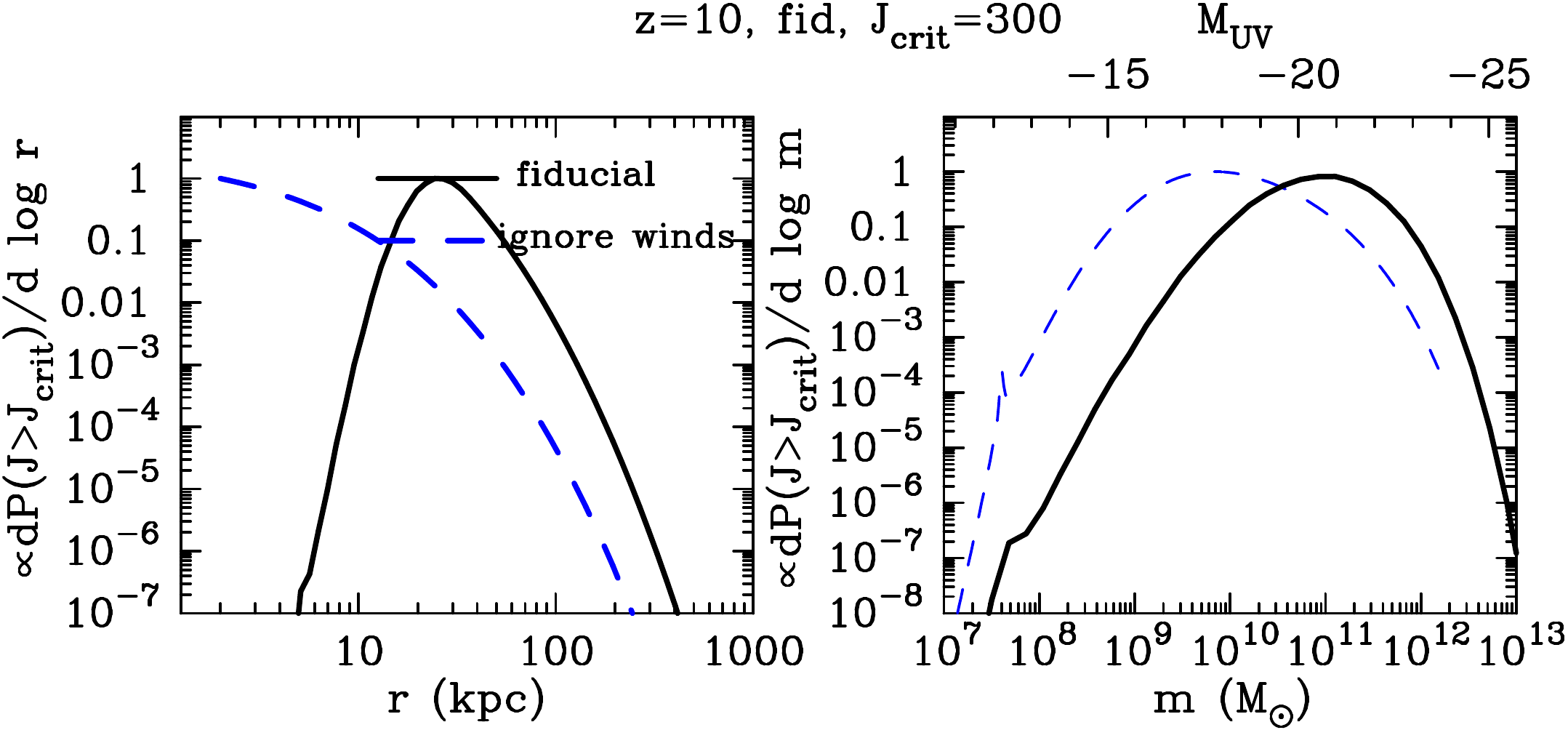,angle=270,width=14.5truecm}}}
\caption{{\it Left panel:}  [{\it right:}] Probability distribution (normalized to peak probability) of the location [mass] of the nearby halo that is exposing a collapsing gas cloud to a $\jlw > \jcrit=300$ at $z=10$. {\it Black solid lines} represent fiducial model (i), while the {\it blue dashed lines} represent model (iv) in which we ignore winds. In the fiducial model, collapsing gas clouds at $z=10$ exposed to $\jlw>\jcrit=300$ have a nearby ($r\gsim 30$ kpc) ultra-luminous star forming galaxy embedded in a dark matter halo of mass $M\gsim 10^{12} M_{\odot}$ with absolute UV magnitude of $M_{\rm UV} \lsim -23$ (as indicated by the upper horizontal axis). If we ignore metals, then it is possible to meet the requirements for DCBHs in closer, fainter star forming galaxies, which reside in less massive halos.}
\label{fig:pdf2}
\end{figure*}

\subsection{Lyman-Werner flux PDF}
\label{sec:fluxpdf}
We introduce the probability that the putative DCBH host halo has an illuminating halo  of mass in the range $M \pm dM/2$ at a separation within the range $r \pm dr/2$:
\begin{equation}
\frac{d^2P}{d M \hs dr}=4 \pi r^2 (1+z)^3[1+\xi]\frac{dn_{\rm ST}}{d M}.
\label{eq:prob}
\end{equation}

Many useful PDFs can be derived from this quantity. For example, if we assume that the LW flux that permeates a collapsing cloud is dominated by a single source, then the probability that the flux is in the range $\log \jlw \pm d \log \jlw /2$ equals 
\begin{eqnarray}
\frac{dP}{d\log \jlw}=\int_{r_{\rm min}}^{r_{\rm max}} dr \\ \nonumber
 \int_{m_{\rm min}}^{\infty} dM\hs \frac{d^2P}{d M \hs dr}\hs\frac{dP(M)}{d\log L_{\rm LW}(r)}\Theta(r-r_{\rm s}[M])
\label{eq:jpdf}
\end{eqnarray} where $L_{\rm LW}(r)=16 \pi^2 r^2\jlw$, and $dP(M)/d\log L_{\rm LW}(r)$ denotes the differential probability that a dark matter halo of mass $M$ has this luminosity $L_{\rm LW}(r)$.  The function $\Theta(x)$ denotes the Heaviside step function (as used in Eq~\ref{eq:mc}), and $r_{\rm s}[M]$ denotes the distance the supernova-driven outflow has traveled after $t=t_{\rm ff}$ (see Eq.~\ref{eq5}). The Heaviside step function ensures that we do not include halos which lie inside the radius of pollution by the neighboring halo/galaxy (i.e. $r>r_{\rm s}$, where $r_{\rm s}$ is given by Eq~\ref{eq4}).

The {\it left panel} of Figure~\ref{fig:pdf1} shows an example $\jlw$-PDFs for our fiducial model at $z=10$ obtained from our Monte-Carlo simulations as {\it open circles}, while the {\it black solid line} shows the PDF given by Eq~\ref{eq:jpdf}. The analytic solution clearly provides a good description of the $\jlw$-PDF at high $\jlw$ (as was shown previously in Dijkstra et al. 2008). For comparison, we have also shown an analytic calculation in which we ignore the winds ({\it blue dashed lines}, i.e. in which we do not include the Heaviside step function in Eq.~\ref{eq:jpdf}), and in which we increased the wind radius by a factor of $2$ ({\it red dot-dashed line}, i.e. in which $r_{\rm s} \rightarrow 2r_{\rm s}$. This comparison shows that metals affect the tail-end of the $\jlw$-PDFs.

The {\it right panel} of Figure~\ref{fig:pdf1} shows the cumulative fraction of collapsing clouds that see a boost $x$ in their LW flux compared to background value. From this we conclude that only $\sim 10^{-8}$ of all halos `see' a $\jlw$ that is $\sim 50$ times $\jbg$, which corresponds to the boost that is required to reach $\jcrit=300$ (represented by the {\it grey region} in both panels). 

\subsection{Pair separation and companion mass PDFs}
Additional interesting PDFs can be computed from Eq~\ref{eq:prob}, e.g., the distribution of positions of the nearby dark matter halo for gas clouds exposed to $\jlw > \jcrit$. This probability is given by
\begin{eqnarray}
\frac{dP(\jlw>\jcrit)}{d\log r}=(\ln 10) r \times \\ \nonumber
 \times\int_{m_{\rm min}}^{\infty}dM \frac{d^2P}{d M \hs dr}P(L_{\rm LW}[M]> L_{\rm crit}[r])\times \Theta(r-r_{\rm m}[m]),
\end{eqnarray} 
where $L_{\rm crit}[r] =16 \pi^2 r^2\jcrit$, and where $P(L_{\rm LW}[M]> L_{\rm crit}[r])$ denotes the probability that a halo of mass $M$ has a LW-luminosity exceeding $L_{\rm crit}[r]$. Examples of these distributions are shown in the {\it left panel} of Figure~\ref{fig:pdf2} for the fiducial model (i) at $z=10$ as {\it black solid lines}, and for model (iv, in which we ignore winds) as {\it blue dashed lines}. This Figure shows that accounting for metal pollution eliminates very close pairs, with $r \lsim 30$ kpc of halos as possible DCBH sites (this number is specific for this particular model). 

Similarly, the distribution of masses of the nearby dark matter halo, if the halo gas is exposed to a $J_{\rm 21} > \jcrit$ equals
\begin{eqnarray}
\frac{dP(\jlw>\jcrit)}{d\log M}= (\ln 10) M \times \\ \nonumber
\times \int_{r_{\rm min}}^{\infty}d r \frac{d^2P}{d M \hs dr} P(L_{\rm LW}[M]> L_{\rm crit}[r])\Theta(r-r_{\rm m}[M]).
\end{eqnarray} 
Examples of this distribution are shown in the {\it right panel} of Figure~\ref{fig:pdf2}. This comparison shows that winds eliminate all galaxies except the most luminous ones ($M_{\rm UV}\sim -21 \pm 1$) populating the most massive halos ($M \gsim 10^{11}M_{\odot}$) as creating the conditions for DCBH formation (also see Fig~\ref{fig:minmass}). This explains why our predicted $n_{\rm DCBH}(z)$ is identical for model (i) and model (ii, in which we truncate the UV luminosity function at $M_{\rm UV}=-14$) (see Fig~\ref{fig:ncan}). This preferred elimination of fainter galaxies embedded within low mass halos is mostly a consequence of the fact that metals eliminate the closest pairs of halos as possible DCBH sites.  

\label{lastpage} 
\end{document}